\PassOptionsToPackage{tbtags}{amsmath}
\documentclass[aip,jcp,amsmath, amssymb, reprint, 
]{revtex4-2}

\usepackage{dcolumn, bm, indentfirst,enumerate}
\usepackage{graphicx, tikz, orcidlink}

\usepackage{hyperref}
\usepackage{xcolor}
\definecolor{AIPBlue}{RGB}{61, 180, 229}
\hypersetup{
 colorlinks,
 linkcolor = AIPBlue,
 citecolor = AIPBlue,
 urlcolor = AIPBlue
}
\newcommand{\D}{{\rm{d}}}

\begin{document}
\title{Coherent two-dimensional THz magnetic resonance spectroscopies for molecular magnets: Analysis of Dzyaloshinskii--Moriya interaction}

\date{Last updated: \today}

\author{Jiaji Zhang \orcidlink{0000-0003-2978-274X}}
\email[Authors to whom correspondence should be addressed: ]{zhang.jiaji.i92@kyoto-u.jp and tanimura.yoshitaka.5w@kyoto-u.jp}

\author{Yoshitaka Tanimura \orcidlink{0000-0002-7913-054X}}
\email[Authors to whom correspondence should be addressed: ]{zhang.jiaji.i92@kyoto-u.jp and tanimura.yoshitaka.5w@kyoto-u.jp}

\affiliation{Department of Chemistry, Graduate School of Science, Kyoto University, Kyoto 606-8502, Japan}

\begin{abstract}
To investigate the novel quantum dynamic behaviors of magnetic materials that arise from complex spin--spin interactions, it is necessary to probe the magnetic response at a speed greater than the spin-relaxation and dephasing processes. Recently developed two-dimensional (2D) terahertz magnetic resonance (THz-MR) spectroscopy techniques use the magnetic components of laser pulses, and this allows investigation of the details of the ultrafast dynamics of spin systems. For such investigations, quantum treatment---not only of the spin system itself but also of the environment surrounding the spin system---is important. In our method, based on the theory of multidimensional optical spectroscopy, we formulate nonlinear THz-MR spectra using an approach based on the numerically rigorous hierarchical equations of motion. We conduct numerical calculations of both linear (1D) and 2D THz-MR spectra for a linear chiral spin chain. The pitch and direction of chirality (clockwise or anticlockwise) are determined by the strength and sign of the Dzyaloshinskii--Moriya interaction (DMI). We show that not only the strength but also the sign of the DMI can be evaluated through the use of 2D THz-MR spectroscopic measurements, while 1D measurements allow us to determine only the strength.
\end{abstract}

\maketitle

\section{Introduction}
\label{sec.intro}
Electron paramagnetic resonance (EPR) and nuclear magnetic resonance (NMR) have a long history as typical examples of two-dimensional (2D) spectroscopy techniques that measure the varying time intervals of magnetic pulse trains applied to electron or nuclear spin systems. \cite{SchramlBook1988, LevittBook2013} Although these techniques are powerful means for structural analysis of organic and inorganic materials, it is difficult to apply them to the investigation of spin dynamics because the time resolution of the magnetic pulses involved is limited to the order of microseconds.

Recently, coherent terahertz (THz) magnetic resonance (MR) spectroscopy was developed using magnetic pulses in the subpicosecond range generated by the magnetic-field component of THz light. \cite{Sato2016PRB, Neu2018JAP,Tanaka2023} Just as ultrafast laser spectroscopy has made it possible to study the electronic excitation-state dynamics and intra- and intermolecular vibrational motions of complex molecular systems, \cite{MukamelBook1995} THz-MR spectroscopy has opened up the possibility of investigating strongly correlated spin dynamics in molecular magnets; this depends on the complex spin--spin interactions and the configurations of spins in the molecular environment, which cause dephasing and relaxation of the spin system.

The quasi-ferromagnetic (FM) and antiferromagnetic (AFM) precession modes in $\rm{YFeO_{3}}$ (YFO) that arise from the antisymmetric spin--orbit coupling \cite{Herrmann1963JPCS} have been observed as free induction decay (FID) signals. \cite{Nakajima2010OE, Yamaguchi2010PRL} Spin-wave excitation in AFM $\rm{NiO}$ has been detected with the aid of the Faraday effect. \cite{Kampfrath2011NP, Sato2010PRL} The skyrmion, which is a quasiparticle composed of vortex-like spin orientations, has been detected in several FM and AFM materials with thin-film structures. \cite{Turgut2017PRB, Dou2023ACSN, Soumyanarayanan2017NM} THz electron spin resonance (ESR) and Hall conductivity spectroscopy have been employed to investigate the topological Hall effect and phase transitions. \cite{Ozerov2014PRL, Hahashi2021NC, Sato2016PRB, Tsuruta2018PRB}

Although THz-MR and THz-ESR spectroscopic approaches based on linear response theory have been successful for the classification of complex spin states in condensed phases, \cite{Biesner2022PRB, Peedu2022PRB, Kozuki2011OE} it is unclear whether the spin states that are investigated are quantum-mechanically entangled, and whether the width of the peaks that are measured arise from the relaxation or dephasing processes, because these peaks are usually broad and overlap. Thus, an extension to coherent 2D THz-MR spectroscopy was developed. The observed 2D THz-MR spectrum for a YFO crystal made it possible to illustrate characteristic nonlinear spin responses, such as double-quantum (2Q) coherence and second-harmonic generation (SHG). \cite{Lu2017PRL}

As already illustrated in 2D optical spectroscopy, \cite{Hamm2011ConceptsAM,Cho2009} 2D THz-MR spectroscopy is not only useful for identifying complex spin interactions but also for monitoring the complex quantum spin dynamics under the influence of relaxation and dephasing arising from the environment at the femtosecond scale. Theoretical input regarding the complex profiles of spin--spin interactions is important for analyzing these 2D spectra under ultrafast nonlinear processes. In particular, this includes those in molecular magnets, which play a central role in spintronics and next-generation information technologies. \cite{Fert2017NRM, Jungwirth2016NN, Inoue2021CL}

In this paper, we provide a comprehensive theoretical framework for both 1D and 2D THz-MR measurements based on the response-function theory developed for nonlinear optical spectroscopy techniques. To illustrate our approach more closely, we employ a chiral spin chain described as a Heisenberg model with exchange coupling and Dzyaloshinskii--Moriya interaction (DMI) arising from the antisymmetric spin--orbit coupling. \cite{TogawaKishineJPSJ2016, KatoKishinePRB2022, Kishine2005PTPS} Such anisotropy of the spin system leads to a series of unique magnetic and optical properties. \cite{Lodahl2017Nature} Unlike with conventional FM materials, in addition to the magnetic anisotropy, a non-centrosymmetric structure arises as the result of the chiral ordering of spins. As a result, a series of new phenomena arising from the nonlinear magnetic response, such as magneto-chiral dichroism and magnetization-induced SHG, were observed. \cite{Mito2009PRB, Okamura2015PRL, Clements2017SR, Nomura2019PRL}

To investigate the properties of such materials as devices, it is necessary to investigate ultrafast spin dynamics under time-dependent external fields, in which quantum coherence and entanglement---not only among spins but also between spins and the environment---play a significant role. \cite{LevittBook2013,DT10PRL} Thus, we include a harmonic heat bath of finite temperature in the spin system. The number of degrees of freedom of the heat bath is then reduced to obtain a time-irreversible equation of motion describing the effects of thermal fluctuations and dissipation. \cite{WeissBook2012, BreuerBook2007} Because the motion of the spins is much faster than the thermal noise arising from the heat bath, the heat bath must be treated in a non-Markovian manner. Thus, the reduced equations of motion derived using Markovian approximations and other assumptions---such as the Bloch, Lindblad, and Redfield equations---are not suitable for the description of such ultrafast spin dynamics. We then employ the numerically ``exact'' hierarchical equations of motion (HEOM) approach, which can be used to treat non-Markovian and non-perturbative system--bath interactions at finite temperatures. \cite{TK89JPSJ1,IT05JPSJ,T06JPSJ,T20JCP}

To demonstrate the applicability of the present theory, 1D and 2D THz-MR spectra were calculated for the chiral spin model with different DMI strengths describing the pitch and direction of chirality (clockwise or counterclockwise). We show that in 1D spectra, only the absolute strength of the DMI can be evaluated, whereas in 2D spectra, the sign of the DMI, which determines the direction of the chirality, can also be determined. While neutron-scatting techniques have been used to determine the structures of chiral materials, the present results indicate the possibility of determining the DMI through spin-dynamic processes. This finding should be valuable for the design of spintronic materials.

The rest of this paper is organized as follows. In Sec.~\ref{sec.theory.thzmr}, we formulate linear (1D) and 2D THz-MR spectra based on the response-function theory. In Sec.~\ref{sec.theory.model}, we introduce the Heisenberg model with the DMI coupled to the harmonic heat bath. The HEOM are then presented. Numerical results and some discussion are presented in Sec.~\ref{sec.result}. Finally, Sec.~\ref{sec.conclude} is devoted to our conclusions.

\section{Magnetic susceptibilities in 2D THz-MR spectroscopy}
\label{sec.theory.thzmr}
We consider a spin system coupled to a bath system driven by an external magnetic field $B(s)$. The total Hamiltonian is expressed as
\begin{equation}
\hat{H}^{\prime}(s) = \hat{H}_{tot} + B(s) \, \hat{M},
\label{eq.thzmr.time_hamiltonian_def}
\end{equation}
where $\hat{H}_{tot}$ is the Hamiltonian of a composite system and $\hat{M}$ is the polarization operator for magnetic fields.

The observable in a magnetic measurement at time $t$ is expressed as $M(t)\equiv\langle \hat{M}(t) \rangle-\langle \hat{M} \rangle$, where $\hat{M}(t)$ is the Heisenberg representation of $\hat{M}$ for $\hat{H}^{\prime}(s)$. The thermal average for any operator $\hat{A}$ is defined as $\langle \hat{A} \rangle \equiv {\rm{tr}}\{\hat{A} \hat{\rho}_{tot}^{eq}\}$, with the equilibrium density operator expressed as $\hat{\rho}_{tot}^{eq}$. We can also express this as ${M}(t)= {\rm{tr}}\{\hat{M} \mathcal{G}'(t)\hat{\rho}_{tot}^{eq}\}- {\rm{tr}}\{\hat{M} \hat{\rho}_{tot}^{eq}\}$, where the Liouville operator is defined as
\begin{align}
\mathcal{G}'(t) \hat{\rho}_{tot}(0) \equiv{\underleftarrow \exp }
\left[- \frac{i}{\hbar}\int_{0}^{t} {\D}s^{\prime} \hat{H}^{\prime}(s^{\prime}) \right]
\hat{\rho}_{tot}(0) \nonumber \\
 \times {\underrightarrow \exp} \left[ \frac{i}{\hbar}
\int_{0}^{t} {\D}s^{\prime}\hat{H}^{\prime}(s^{\prime})\right].
\end{align}
Here, the arrows indicate time-ordered exponentials.

Experiments such as 2D NMR and 2D EPS measurements \cite{SchramlBook1988,LevittBook2013} are conducted using multiple magnetic pulses with finite time widths. In such experiments, the excitation by the external field is non-perturbative, and the desired spin dynamics are investigated by designing the profiles of pulse trains, as in the cases of spin-echo and correlation spectroscopy measurements. Theoretically, these signals are obtained by integrating the equations of motion for a spin system, such as the Bloch, \cite{Bloch1946PR} Redfield, \cite{Redfield1965AMOR} or stochastic Liouville equations, \cite{KuboBook1969} or the HEOM, \cite{T06JPSJ, JT08CPL, TT20JPSJ} under a sequence of magnetic pulses.

As in the case of coherent optical laser spectroscopies, the excitations of coherent THz-MR spectroscopy are assumed to be impulsive, which allows us to measure the signals in different orders of the field-system interactions separately. Thus, we can employ a response-function theory developed for ultrafast nonlinear laser spectroscopy. \cite{MukamelBook1995} Up to the third order, the signal is then expressed as
\begin{equation}
\begin{split}
M(t) &= \int_{0}^{t} {\D}s \, \chi_{1}(s) \, B(t-s) \\
&+ \int_{0}^{t} {\D}s_{1} \int_{0}^{s_1} {\D}s_{2} \,
\chi_{2}(s_1, s_2) \, B(t- s_{1} ) \, B(t- s_{2}) \\
&+ \int_{0}^{t} {\D}s_{1} \int_{0}^{s_1} {\D}s_{2} \,
\int_{0}^{s_2} {\D}s_3 \, \chi_{3}(s_1,s_2, s_3) \, \\
&~~~~~~~~~~~~~~~~\times B(t- S_1) \, B(t- s_{2}) \,
B(t- s_{3}),
\end{split}
\label{eq.thzmr.third_order}
\end{equation}
where the linear, second-order, and third-order response functions are defined as \cite{T06JPSJ}
\begin{equation}
\begin{split}
\chi_{1}(s) &\equiv { -\frac{i}{\hbar}}\langle[ \hat{M}(s), \hat{M} ]\rangle \\
&={ -\frac{i}{\hbar}} {\rm{tr}}\left\{ \hat{M} \mathcal{G}(s) \,
\hat{M}^{\times} \hat{\rho}_{tot}^{eq} \right\},
\label{eq.thzmr.linear_resp_def}
\end{split}
\end{equation}
\begin{equation}
\begin{split}
\chi_{2}(s_1, s_2) &\equiv { -\frac{1}{\hbar^2} } \langle[ [\hat{M}(s_1+s_2),
\hat{M}(s_1)], \, \hat{M} ]\rangle \\
&= { -\frac{1}{\hbar^2} } {\rm{tr}}\left\{ \hat{M} \, \mathcal{G}(s_2) \,
\hat{M}^{\times} \mathcal{G}(s_1)\hat{M}^{\times} \hat{\rho}_{tot}^{eq} \right\},
\label{eq.thzmr.second_resp_def}
\end{split}
\end{equation}
and
\begin{equation}
\begin{split}
\chi_{3}&(s_1, s_2, s_3) \\
&\equiv { \frac{i}{\hbar^3} } \langle
[[[\hat{M}(s_1+s_2+s_3), \hat{M}(s_1+s_2)], \, \hat{M}(s_1) ], \, \hat{M}] \rangle \\
&= { \frac{i}{\hbar^3} } {\rm{tr}}
\left\{ \hat{M} \, \mathcal{G}(s_3)\, \hat{M}^{\times} \, \mathcal{G}(s_2) \,
\hat{M}^{\times} \mathcal{G}(s_1)\hat{M}^{\times} \hat{\rho}_{tot}^{eq} \right\}.
\label{eq.thzmr.third_resp_def}
\end{split}
\end{equation}
Here, $\mathcal{G}(s)$ is the Liouvillian without magnetic interaction, which is defined from $\mathcal{G}'(s)$ with $B(s)=0$, and we have introduced the superoperator $\hat{M}^{\times }{{\hat \rho}}\equiv [\hat{M},{\hat \rho}]$. The right-hand side (rhs) of the second line in each of the above equations allows us to employ the equations of motion to calculate the response functions and give us an intuitive picture of the higher-order optical processes. \cite{T06JPSJ,T20JCP} For example, the rhs of the second line of Eq.~\eqref{eq.thzmr.linear_resp_def} can be read from right to left as follows. The total system is initially in the equilibrium state $\hat{\rho}_{tot}^{eq}$. The initial state is then modified by the first magnetic pulses via the dipole operator as $(\hat{M}^{\times} \hat{\rho}_{tot}^{eq})=[\hat{M}, \hat{\rho}_{tot}^{eq}]$ at $t = 0$ and is propagated for time $t=s$ by $ \mathcal{G}(s)$. The expectation value is then obtained by calculating the trace of $\hat M$.

Actual 2D experiments have been conducted using a pair of magnetic pulses $a$ and $b$ with inter-pulse delay $\tau$. \cite{Lu2017PRL, Wan2019PRL} Under the impulsive approximation, the magnetic field is expressed as
\begin{equation}
B(s) = B_{a} \delta(s) + B_{b} \delta(s-\tau) ,
\label{eq.thzmr.magnetic_field}
\end{equation}
where $B_{a}$ and $B_{b}$ are the magnetic field strengths. The nonlinear element of the signal $M_{NL}(t,\tau)$ at $t+\tau$ is then evaluated as
\begin{equation}
M_{NL}(t,\tau) = M_{ab}(t+\tau) - M_{a}(t+\tau) - M_{b}(t+\tau),
\label{eq.thzmr.nonlinear_signal_def}
\end{equation}
where $M_{ab}(t+\tau)$, $ M_{a}(t)$, and $ M_{b}(t)$ are the total signal and the linear elements for pulses $a$ and $b$, respectively.

From Eqs.~\eqref{eq.thzmr.third_order} and \eqref{eq.thzmr.magnetic_field}, the nonlinear element is evaluated as
\begin{equation}
\begin{split}
M_{NL}(t, \tau) &= B_{a} B_{b} \, \chi_{2}(\tau, t) +
B_{a} B_{b}^{2} \, \chi_{3}(\tau, 0, t) \\
&+ B_{a}^2 B_{b} \, \chi_{3}(0, \tau, t).
\end{split}
\label{eq.thzmr.nonlinear_signal_response}
\end{equation}
Thus, the characteristic features of THz-MR spectroscopy are described by the nonlinear susceptibilities. Because each term on the rhs has different proportionality with respect to $B_a$ and $B_b$, they can be evaluated separately by changing their respective field strengths.

In nonlinear optical spectroscopies, $\chi_{2}$ is used to analyze 2D THz-Raman \cite{HammPerspH2O2017,IIT15JCP} and 2D THz-IR-Raman signals \cite{NagataChemRev2016,TT23JCP1} that involve a 2Q transition process, while $\chi_{3}(s_3, s_2, s_1)$ is used to analyze nonlinear 2D THz spectroscopies \cite{Reimann2021JCP,OT98CPL} including 2D THz rotational spectroscopy. \cite{Nelson2016PNAS,IT19JCP} As illustrated in nonlinear optical spectroscopies, the $s_1$ and $s_3$ periods describe the time evolutions of coherent states, while $s_2$ describes the evolution of population states. \cite{Hamm2011ConceptsAM,Cho2009} While $\chi_{3}(\tau, 0, t)$ is used for 2D spectroscopies with waiting time $s_2=0$, $\chi_{3}(0, \tau, t)$ is used for transient absorption spectra in which the time evolution of the excited-state population is measured.

Using $\chi_{1}(s)$ in Eq.~\eqref{eq.thzmr.linear_resp_def}, the linear absorption spectrum is defined as
\begin{equation}
\chi_{1}(\omega) = {\rm{Im}} \int_{0}^{\infty} {\D}s e^{- i \omega s} \chi_{1}(s),
\label{eq.thzmr.spectrum_linear}
\end{equation}
where {\rm{Im}} denotes the imaginary part. Following the experimental setup, we consider two kinds of 2D spectrum for different time configurations of $\tau$ and $t$ in Eq.~\eqref{eq.thzmr.third_resp_def}, expressed as: (1)~$\chi_{3}(\tau, 0, t)$ and (2)~$\chi_{3}(0, \tau, t)$. They are then expressed in the Fourier translation form as
\begin{equation}
\chi_{3}^{(1)}(\omega_\tau, \omega_t) = {\rm{Im}} \int_{0}^{\infty}
{\D}\tau \int_{o}^{\infty} {\D}t \, e^{-i \omega_\tau \tau - i \omega_t t}
\chi_{3}(\tau, 0, t),
\label{eq.thzmr.spectrum_2D_0}
\end{equation}
\begin{equation}
\chi_{3}^{(2)}(\omega_\tau, \omega_t) = {\rm{Im}} \int_{0}^{\infty}
{\D}\tau \int_{o}^{\infty} {\D}t \, e^{-i \omega_\tau \tau - i \omega_t t}
\chi_{3}(0, \tau, t),
\label{eq.thzmr.spectrum_2D_1}
\end{equation}
where $\omega_\tau$ and $\omega_t$ represent the excitation and detection frequencies, respectively. \cite{Lu2017PRL}

\section{Chiral spin model and HEOM approach}
\label{sec.theory.model}
To describe the novel phenomena of chiral magnets, we consider a linear chain system consisting of $L$ spins. The system Hamiltonian is defined as
\begin{equation}
\begin{split}
\hat{H}_{S} &= J \sum_{n=1}^{L} \left[ \hat{\sigma}_{n}^{z} \,
\hat{\sigma}_{n+1}^{z}+ \Delta \left( \hat{\sigma}_{n}^{x} \,
\hat{\sigma}_{n+1}^{x} + \hat{\sigma}_{n}^{y} \,
\hat{\sigma}_{n+1}^{y} \right) \right] \\
&+ D_{y} \sum_{n=1}^{L} \left[ \hat{\sigma}_{n}^{z} \,
\hat{\sigma}_{n+1}^{x} - \hat{\sigma}_{n}^{x} \,
\hat{\sigma}_{n+1}^{z} \right] ,
\end{split}
\label{eq.theory.system_Hamiltonian}
\end{equation}
where $\hat{\sigma}_{n}^{\alpha}$ $(\alpha = x, y, z)$ denotes the Pauli operator at the $n$-th site in the $\alpha$-th direction. Here, the first term with $J>0$ represents the AFM Heisenberg exchange coupling, and $\Delta$ is the anisotropic parameter. The second term in Eq.~\eqref{eq.theory.system_Hamiltonian} is the DMI that is perpendicular to the $xz$ plane, with coupling strength $D_{y}$. The sign of $D_{y}$ determines the direction (i.e., clockwise or anticlockwise) or handedness (i.e., right- or left-handed circular configuration) of chirality, which is not easily determined by experimental measurements. \cite{Sato2016PRB, Kishine2005PTPS} The magnetization operator is defined as
\begin{equation}
\hat{M} = \sum_{l=1}^{L} \hat{\sigma}_{l}^{x} .
\end{equation}

A very important aspect of investigating ultrafast dynamics in magnetic materials is the inclusion of a quantum-mechanically consistent relaxation and dephasing mechanism. This can be achieved by including a harmonic heat bath in the spin system. The Hamiltonian of this heat bath is defined as
\begin{equation}
\hat{H}_{B} = \sum_{j} \left[ \frac{\hat{p}_{j}^{2}}{2m_{j}}
+ \frac{m_{j}\omega_{j}^{2}}{2} \hat{x}_{j}^{2} \right],
\end{equation}
where $\hat{p}_{j}$, $\hat{x}_{j}$, $\omega_{j}$, and $m_{j}$ represent the momentum, position, frequency, and mass of the $j$-th oscillator, respectively. The system--bath interaction is defined by
\begin{equation}
\hat{H}_{I} = \hat{V} \, \sum_{j} g_{j} \, \hat{x}_{j},
\end{equation}
where $\hat{V}$ represents the spin part of the system--bath interaction function, defined as
\begin{equation}
\hat{V} = \sum_{n}^{L} \hat{\sigma}_{n}^{z},
\end{equation}
and $g_{j}$ is the coupling strength with the $j$-th oscillator. The total Hamiltonian is then given by
\begin{equation}
\hat{H}_{tot} = \hat{H}_{S} + \hat{H}_{I} + \hat{H}_{B} .
\end{equation}

In open quantum dynamics theory, the time-irreversible process can be described using a reduced set of equations of motion. After reducing the number of degrees of freedom of the heat bath, the noise effects are characterized by the correlation function,
\begin{equation}
C(t) = \frac{1}{\pi} \int_{0}^{\infty} {\D}\omega \,
J(\omega) \left[ \coth\left(\frac{\beta\hbar\omega}{2}\right)
\cos(\omega t) - i \sin(\omega t) \right],
\label{eq.theory.corr_func_def}
\end{equation}
where $J(\omega)$ is the spectral density function (SDF), $\beta = 1/k_{B} T$ is the inverse temperature, and $k_{B}$ is the Boltzmann constant. For the heat bath, we assume the Drude SDF, which is expressed as,
\begin{equation}
J(\omega) = \frac{\zeta \gamma^2 \omega}{\gamma^2 + \omega^2},
\end{equation}
where $\zeta$ is the coupling strength and $\gamma$ is the inverse correlation time.

In non-perturbative and non-Markovian conditions, the time evolution can be described by the HEOM approach. \cite{TK89JPSJ1,IT05JPSJ} We rewrite Eq.~\eqref{eq.theory.corr_func_def} in a linear-summation form,
\begin{equation}
C(t) = \sum_{k}^{K} c_{k} \, e^{-\nu_{k} |t|},
\end{equation}
where $c_{k}$ and $\nu_{k}$ are complex-valued coefficients. The HEOM can then be expressed as \cite{T06JPSJ, T20JCP}
\begin{equation}
\begin{split}
\frac{\partial } {\partial t}\hat{\rho}_{[\vec{n}]}(t) = &- \left[ i \hat{H}_{S}^{\times}
+ \sum_{k}^{N} n_{k} \nu_{k} \right] \hat{\rho}_{[\vec{n}]}(t) \\
& -
i \hat{V}^{\times} \sum_{k}^{K} \hat{\rho}_{[\vec{n}+\vec{e}_{k}]}(t) \\
& - i \sum_{k}^{K} \left[ c_{k} \hat{V} \hat{\rho}_{[\vec{n}-\vec{e}_{k}]}(t)
- c_{k}^{\ast} \hat{\rho}_{[\vec{n}- \vec{e}_{k}]}(t) \hat{V}\right],
\end{split}
\label{eq.theory.heom_def}
\end{equation}
where 
$\vec{n} = \{ n_{1}, n_{2}, \cdots, n_{K} \}$ denotes the index vector, and $n_{k}$ are non-negative integers. Among all the $\hat{\rho}_{[\vec{n}]}(t)$, the $0$-th, with $\vec{0} = \{ 0, 0, \cdots, 0 \}$, corresponds to the density operator of the reduced system, while all the others are introduced for ancillary purposes.

In the HEOM approach, the response functions Eqs.~\eqref{eq.thzmr.linear_resp_def}--\eqref{eq.thzmr.third_resp_def} are evaluated as the time evolution of the system under external excitation. The density matrix is replaced by the HEOM elements, and the Liouvillian $\mathcal{G}(t)$ is replaced using Eq.~\eqref{eq.theory.heom_def}.

For example, we can evaluate $\chi_{3}(\tau, 0, t)$ using the expression of the second line in Eq.~\eqref{eq.thzmr.third_resp_def} as follows. \cite{T06JPSJ,T20JCP} We first run the HEOM program for a sufficiently long period from a temporally initial condition at $t=-t_i$ (such as the factorized initial condition, $\hat{\rho}_{[\vec{0}]}(-t_i)=\exp[-\beta \hat{H}_S]$ with all the other hierarchy elements set to zero) to time $t=0$ to reach a true thermal equilibrium, denoted by $\hat{\rho}_{[\vec{n}]}^{eq}(0)$. All the hierarchy elements have to be used to define a correlated initial condition. The system is next excited by the first magnetic interaction $\hat M$ at $t=0$ as $\hat{\rho}_{[\vec{n}]}'(0) =[ \hat M, \hat{\rho}_{[\vec{n}]}^{eq}(0)]$. The perturbed elements $\hat{\rho}_{[\vec{n}]}'$ then evolve in time by numerically integrating Eq.~\eqref{eq.theory.heom_def} up to $t$. At $t$, the system is excited by the second and third magnetic interactions as $\hat{\rho}_{[\vec{n}]}''(t) = [ \hat M, [ \hat M, \hat{\rho}_{[\vec{n}]}'(t)] ]$. Then, after $\hat{\rho}_{[\vec{n}]}''$ evolves in time with the initial condition $\hat{\rho}_{[\vec{n}]}''(t)$ to $t+\tau$, the response function is calculated from the expectation value of the magnetic moment as $\chi_{3}(\tau, 0, t)=tr_{A}\{ \hat M \hat{\rho}_{[\vec{0}]}''(t+\tau)\}$.

Notice that to take the system--bath coherence (or bath entanglement \cite{T20JCP}) into account during the external perturbation, it is important to operate $\hat M$ to all of the hierarchy elements $\hat{\rho}_{[\vec{n}]}''(t)$. Although we only use $\hat{\rho}_{[\vec{0}]}(t)$ to calculate an expectation value, the other elements are essential to obtain an echo signal for non-Markovian noise in 2D spectroscopy. \cite{T06JPSJ,T20JCP}

\section{Numerical Results}
\label{sec.result}
In the following, we set the exchange-coupling strength as the base unit $J=1$ and calculate 1D (or linear absorption) and 2D spectra for
$\Delta = 0.1$ as a model for typical AFM metal oxides. \cite{Braun2005NP, Thoma2022IEEE} The number of spin sites was $L=6$, with periodic boundary conditions $\hat{\sigma}_{n}^{\alpha} = \hat{\sigma}_{n+L}^{\alpha}$. Here, we consider cases without the DMI ($D_{y}=0$) and with the weak DMI ($|D_{y}| \le 0.05$), which is appropriate for a small model system. Although the HEOM used in this study can investigate cases in which the system interacts strongly with the heat bath, here, we focus on the characterization of the chiral spin system and keep the coupling weak. Thus, the bath parameters were chosen as $\zeta = 0.05$, $\gamma = 1.0$, and $\beta\hbar = 10$. The hierarchy parameters were chosen as $K = 50$ and $\sum_{k}^{K} n_{k} \le 10$, and the time step for numerical integration was set to $\delta t = 0.01$.

\subsection{\texorpdfstring{Energy eigenstates and transition elements of $\hat{M}$}{Energy eigenstates and transition elements of M}}
\label{sec.result.energy}

\begin{table}[ht]
\caption{\label{tab:eigen} Eigenenergies of spin Hamiltonian, $\hat{H}_{S}$ (a)~without the DMI and (b)~with the DMI. }
\centering
\begin{tabular}{c|c|c}
\hline
\hline
& (a)~$D_{y}=0$ & (b)~$D_{y}=\pm 0.05$ \\
\hline \hline
$|0_{0}\rangle$ & 0 & 0 \\
\hline
$|0_{1}\rangle$ & 0.001 & 0.001 \\
\hline
{$|1_{0}\rangle$} &
{$1.02$} &
{$0.99$} \\
\hline
{$|1_{0}^{\prime}\rangle$} &
{$-$} &
{$1.14$} \\
\hline
{$|1_{1}\rangle$} &
{$0.87$} &
{$0.81$} \\
\hline
{$|1_{1}^{\prime}\rangle$} &
{$-$} &
{$0.98$} \\
\hline
{$|1_{2}\rangle$} &
{$0.93$} &
{$0.93$} \\
\hline
$|1_{3}\rangle$ & 1.15 & 1.15 \\
\hline
$|2_{0}\rangle$ & 2.02 & 1.99 \\
\hline
$|2_{1}\rangle$ & 2.03 & 2.03 \\
\hline
$|2_{2}\rangle$ & 1.85 & 1.86 \\
\hline
{$|2_{3}\rangle$} &
{$2.02$} &
{$2.02$} \\
\hline
{$|2_{4}\rangle$} &
{$2.12$} &
{$2.12$} \\
\hline
\hline
\end{tabular}
\end{table}

\begin{figure}[ht]
\centering
\includegraphics[width=0.5\textwidth]{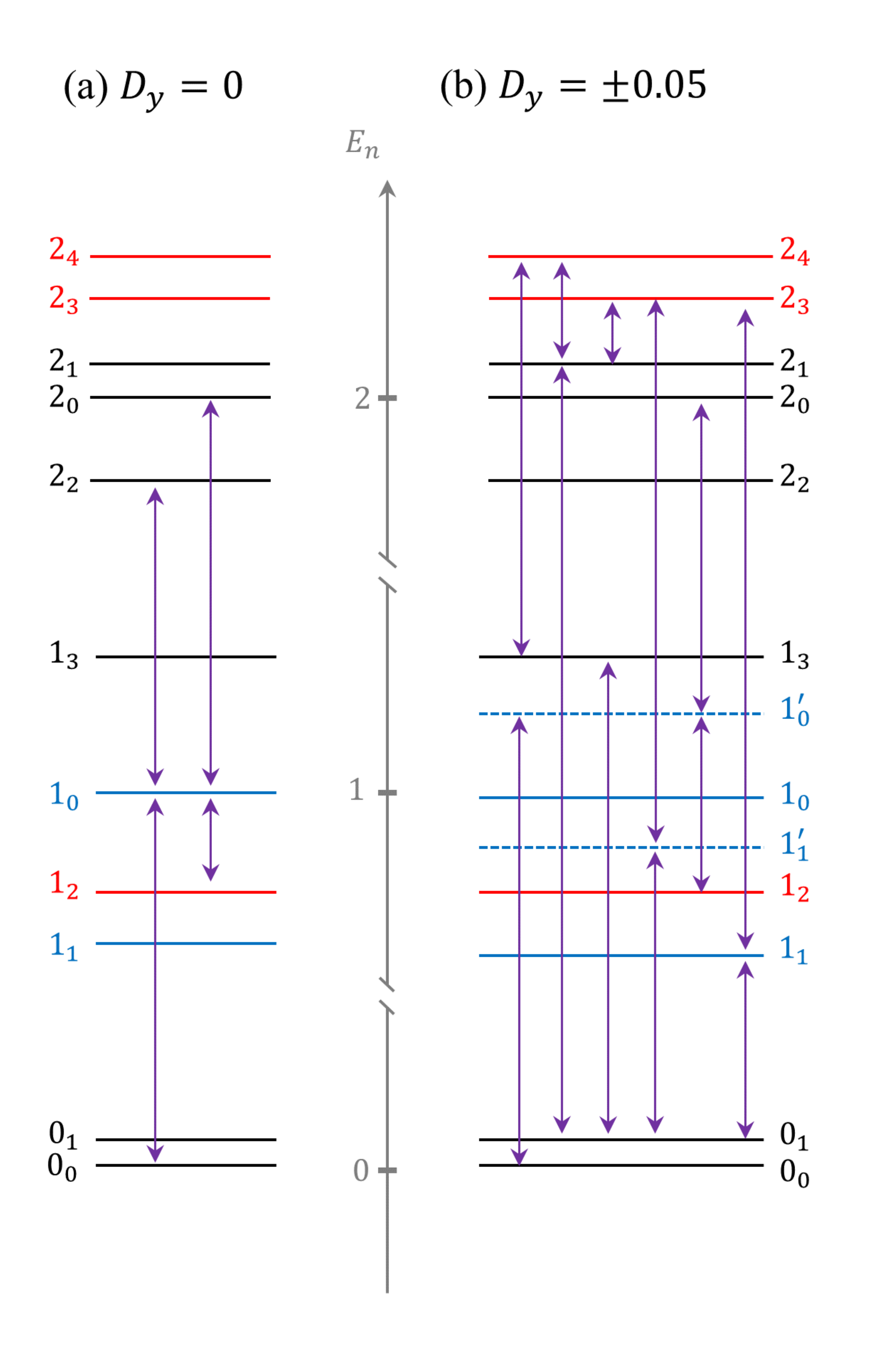}
\caption{Schematic view of the energy states of the spin Hamiltonian, $\hat{H}_{S}$, (a)~without the DMI $(D_{y}=0)$ and (b)~with the DMI $(D_{y}=\pm 0.05)$. The black, blue, and red lines represent the states that are (I)~independent of the DMI, (II)~dependent on the square of $D_{y}$, and (III)~independent of the DMI but the phases of their wave functions change depending on the sign of $D_{y}$, respectively. The purple arrows denote the possible transitions that arise from $\hat{M}$.}
\label{fig.energy_level}
\end{figure}

To illustrate the origin of the peaks in the 1D and 2D spectra that will be shown later, in Table~\ref{tab:eigen}, we present some representative energy eigenstates $|n_{m}^{(\prime)}\rangle$ that are necessary to explain THz-MR spectra. Here, $n = 0$, $1$, and $2$ represent the ground, first, and second excited states, respectively, and $m$ and $^{\prime}$ are introduced to signify the energy splitting arising from the anisotropic coupling $\Delta$ and the DMI, respectively. Thus, for example, the states $|1_{0}\rangle$ and $|1_{0}^{\prime}\rangle$ are degenerate when $D_{y}=0$. Note that to conduct numerical simulations, we employed all of the spin-$z$ basis states to maintain the numerical accuracy.

In Fig.~\ref{fig.energy_level}, we depict the energy eigenstates (a)~without the DMI $(D_{y}=0)$ and (b)~with the DMI $(D_{y} = \pm 0.05)$. The possible magnetic transitions that arise from $\hat{M}$ are denoted by vertical arrows. Here, depending on the role of $D_{y}$, we classify the eigenstates as (I)~independent of the DMI (black lines), (II)~dependent on the square of $D_{y}$ (blue lines), and (III)~independent of the DMI but the phases of their wave functions change depending on the sign of $D_{y}$ (red lines). For a finite $D_{y}$ in Fig.~\ref{fig.energy_level}(b), the degeneracy of eigenenergy is resolved because the inversion symmetry is broken, and we observe the finite energy gap in (II) (blue dashed lines).

\subsection{1D THz-MR spectrum}

\begin{figure}
\centering
\includegraphics[width=0.5\textwidth]{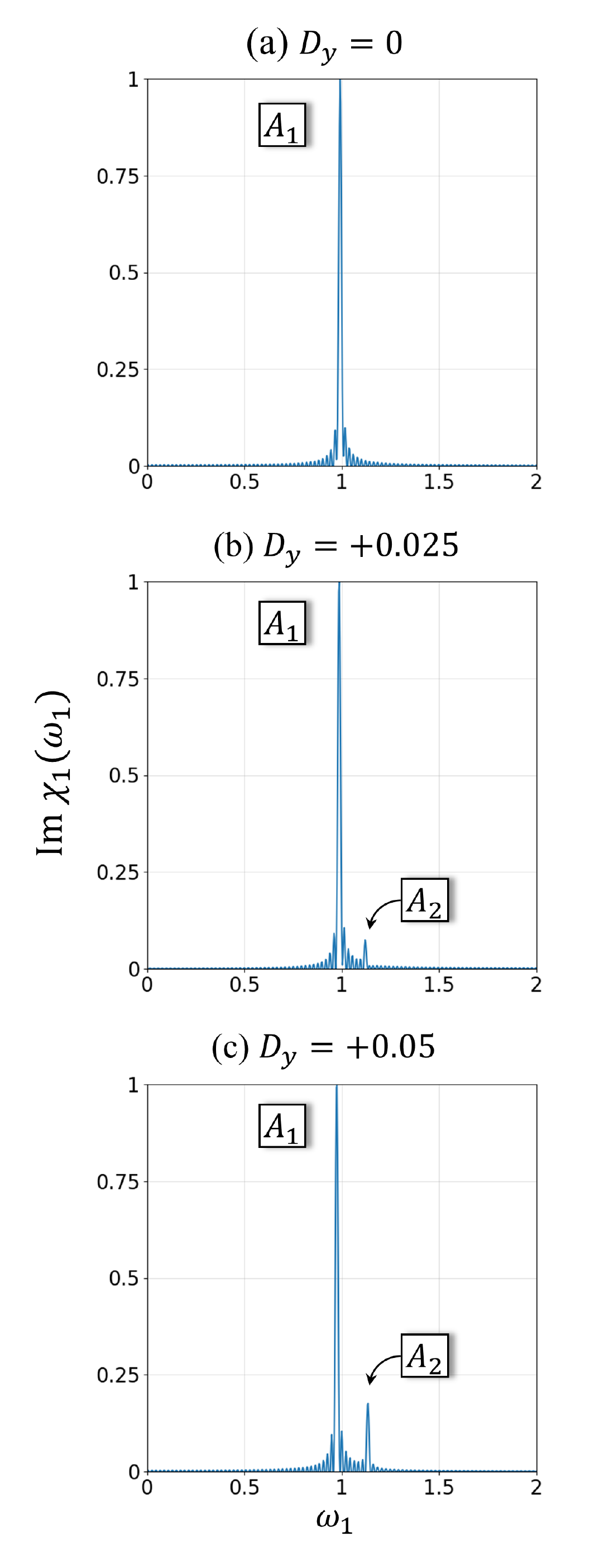}
\caption{Linear absorption (1D) spectra, $\chi_{1}(\omega)$ calculated for $D_{y}= $ (a)~$0$, (b)~$+0.025$, and (c)~$+0.05$. The signal intensities are normalized with respect to the absolute values of the peak intensities.}
\label{fig.linear}
\end{figure}

In Fig.~\ref{fig.linear}, linear absorption (1D) spectra $\chi_{1}(\omega)$ calculated from Eq.~\eqref{eq.thzmr.spectrum_linear} with Eq.~\eqref{eq.thzmr.linear_resp_def} are presented for different $D_{y}$. Here, we only focus on $D_{y}>0$ because the spectral profiles for $-|D_{y}|$ are identical to those for $+|D_{y}|$. The tiny symmetrical peaks around the main peak arise as an artifact of numerical Fourier transformation; these can be suppressed by increasing the time interval and introducing window functions.

Under current low-temperature conditions, the spin system is almost in the ground equilibrium states $|0_{0} \rangle$ and $|0_{1}\rangle$. The populations of other excited states are less than $0.1\%$, and we cannot observe the transitions from the excited states in the 1D spectrum. The main peak $A_{1}$ arises from the transition $|0_{0} \rangle \to |1_{0} \rangle$. Because the DMI resolves the degeneracy of states $|1_{0} \rangle$ and $|1_{1} \rangle$ (blue lines), the energy states $|1_{0}^{\prime} \rangle$ and $|1_{1}^{\prime} \rangle$ (blue dashed lines) appear for finite DMI, as illustrated in Fig.~\ref{fig.energy_level}(b). Accordingly, we observe the adjoint peak $A_{2}$ that arises from the transition $|0_{0} \rangle \to |1_{0}^{\prime}\rangle$. The energy eigenvalues of $|1_{0} \rangle$ and $|1_{0}^{\prime} \rangle$ are (II) dependent upon the square of $D_{y}$, and we found that, for our system with $\Delta /J = 0.1$, these peak positions are evaluated as $\omega_{A_1} \approx -8.34 \, D_{y}^2 + 1.02$ and $\omega_{A_2} \approx 8.34 \, D_{y}^2 + 1.12$. Thus, from the position of the $A_{2}$ peak, we can estimate the magnitude of $D_{y}$, but we cannot determine the direction of chirality, which is described as the sign of $D_{y}$.

\subsection{2D THz-MR spectra}
To elucidate the chirality of the system further, we next present numerical results for 2D THz-MR spectrum evaluated from $\chi_{3}^{(1)}(\omega_\tau, \omega_t)$ and $\chi_{3}^{(2)}(\omega_\tau, \omega_t)$, expressed as Eqs.~\eqref{eq.thzmr.spectrum_2D_0} and \eqref{eq.thzmr.spectrum_2D_1} with Eq.~\eqref{eq.thzmr.third_resp_def}. In the present case, the rephasing parts ($\omega_t<0$ and $\omega_{\tau}>0$) in both 2D spectra have no signal, indicating that there is no spin echo. {\cite{Hamm2011ConceptsAM,Cho2009} Thus, we only present the non-rephasing parts of the spectra ($\omega_t>0$ and $\omega_{\tau}>0$).

\subsubsection{\texorpdfstring{Contribution from $\chi_{3}^{(1)}(\omega_\tau, \omega_t)$}{Contribution from X31(OmegaTau,Omegat)}}

\begin{figure}
\centering
\includegraphics[width=0.5\textwidth]{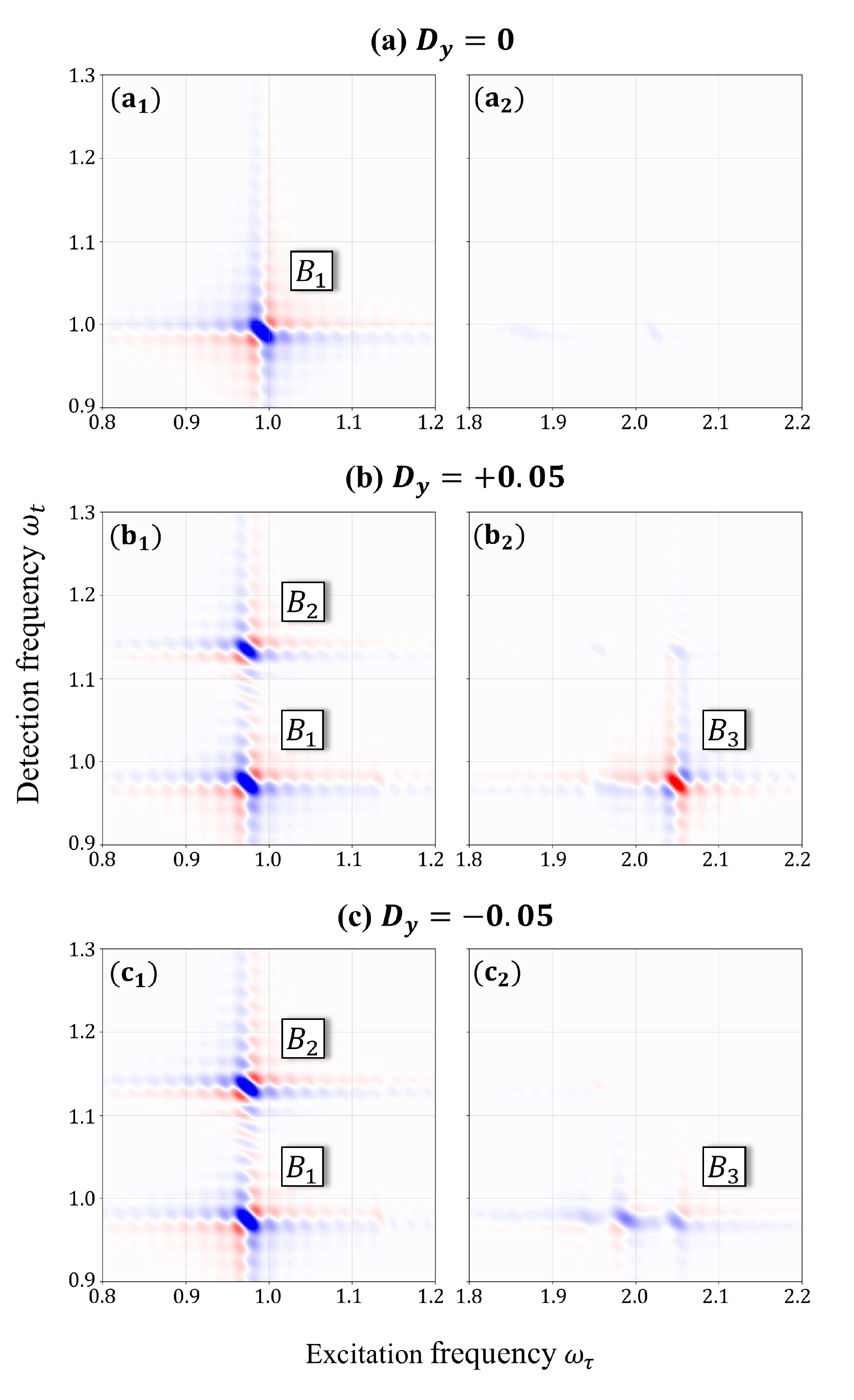}
\caption{2D THz-MR spectrum calculated from $\chi_{3}^{(1)}(\omega_\tau, \omega_t)$ for three values of $D_{y}$: (a)~$0$, (b)~$+0.05$, and (c)~$-0.05$. The intensity of each spectrum is normalized with respect to its maximum peak intensity. The red and blue areas represent the positive and negative intensities of the spectra. }
\label{fig.2D_t0t}
\end{figure}

\begin{figure}
\centering
\includegraphics[width=0.5\textwidth]{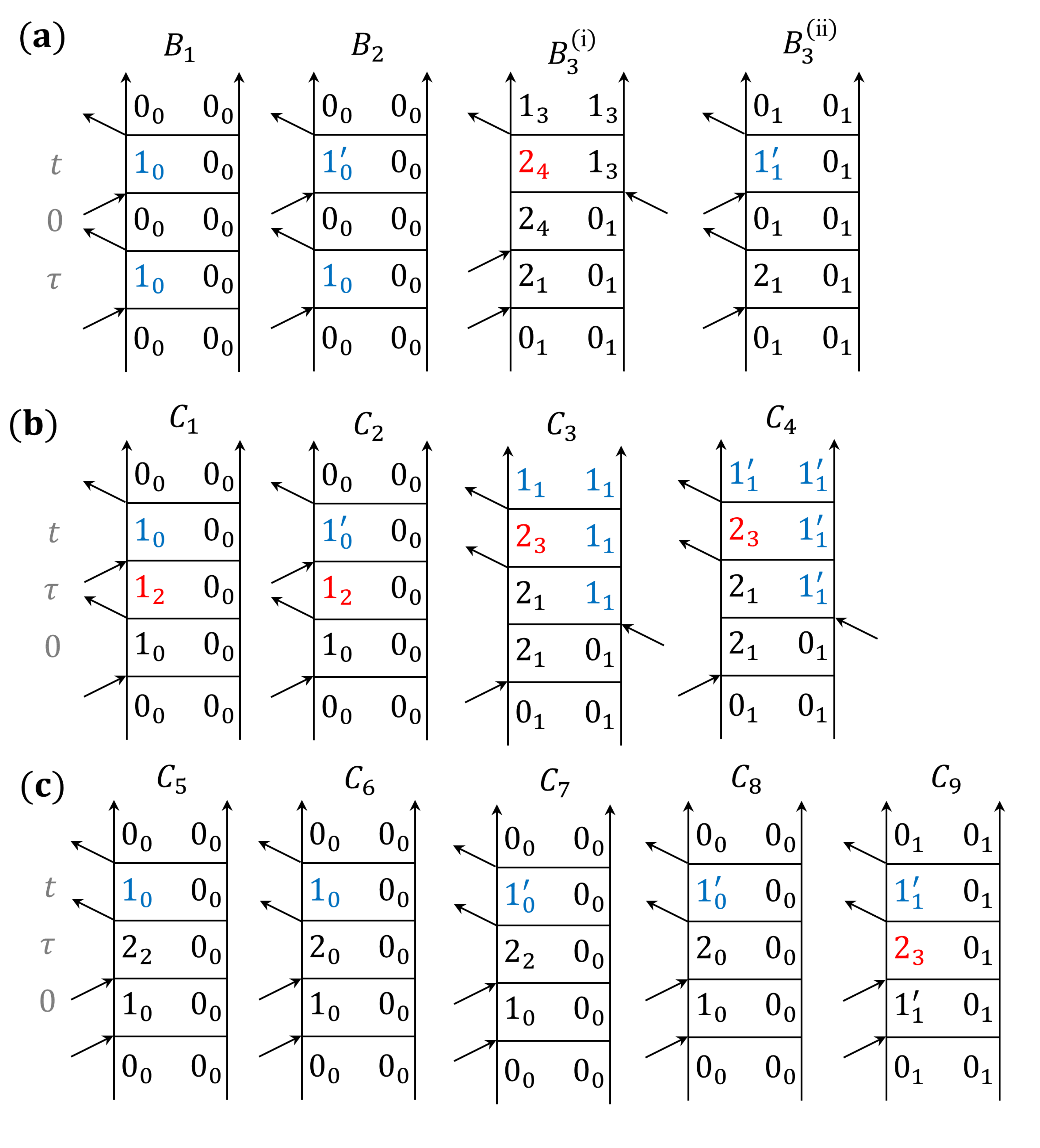}
\caption{Double-sided Feynman diagrams for all labeled peaks. Two different contributions of $B_{3}$ are shown separately.}
\label{fig.2D_paths}
\end{figure}
In Fig.~\ref{fig.2D_t0t}, we present contour maps of $\chi_{3}^{(1)}(\omega_\tau, \omega_t)$ in the (a)~$D_{y}=0$, (b)~$+0.05$, and (c)~$-0.05$ cases, in which the red and blue areas represent positive and negative intensities. To analyze the physical origin of each peak, we present a double-sided Feynman diagram corresponding to each labeled peak in Fig.~\ref{fig.2D_paths}(a). Here, the $B_{3}$ peak consists of two Liouville paths, which we denote as $B_{3}^{(\romannumeral1)}$ and $B_{3}^{(\romannumeral2)}$.

These diagrams describe the time evolution of the left and right wave functions of the density operators involved in the response function from bottom to top under the magnetic excitations and deexcitations. For example, diagram $B_{3}^{(\romannumeral1)}$ in Fig.~\ref{fig.2D_paths}(a) describes the time evolution of the density operator undergoing magnetic interactions described by the purple arrows in Fig.~\ref{fig.energy_level} with the left wave function at time $s=0$, $s=\tau$, and $s=\tau+t$ with the right wave function at time $s=\tau$. The left wave function evolves with time $|0_{1} \rangle \rightarrow |2_{1} \rangle \rightarrow |2_{4} \rangle \rightarrow |1_{3} \rangle $, while the right one evolves with time $\langle 0_{1} | \rightarrow \langle 1_{3} |$. \cite{MukamelBook1995, T20JCP}

The peaks labeled as $B_{1}$ and $B_{2}$ along $\omega_t=0.98$ correspond to peaks $A_{1}$ and $A_{2}$ in the 1D spectra, and the frequency difference $\omega_{A_2} - \omega_{A_1}$ is equal to $\omega_{B_2} - \omega_{B_1}$. As illustrated in Fig.~\ref{fig.2D_paths}(a), the diagrams $B_{1}$ and $B_{2}$ only involve the eigenstates that are (I)~independent of the DMI and (II)~dependent on the square of $D_{y}$. Thus, the intensities of the $B_{1}$ and $B_{2}$ peaks do not depend on the sign of $D_{y}$.

However, the $B_{3}^{(\romannumeral2)}$ diagram in Fig.~\ref{fig.2D_paths}(a) involves states (II) and (III), while the $B_{3}^{(\romannumeral1)}$ diagram contains the state $|2_{4} \rangle $ that is (III)~independent of the DMI but the phase of its wave function changes depending on the sign of $D_{y}$. As a result, the sign of the peak intensity changes for $D_{y} = +0.05$ and $-0.05$. However, the peak profiles in Figs.~\ref{fig.2D_t0t}(b$_2$) and \ref{fig.2D_t0t}(c$_2$) are not positively and negatively symmetric due to the contribution from $B_{3}^{(\romannumeral2)}$.

\subsubsection{\texorpdfstring{Contribution from $\chi_{3}^{(2)}(\omega_\tau, \omega_t)$}{Contribution from X32(OmegaTau,Omegat)}}

\begin{figure}
\centering
\includegraphics[width=0.5\textwidth]{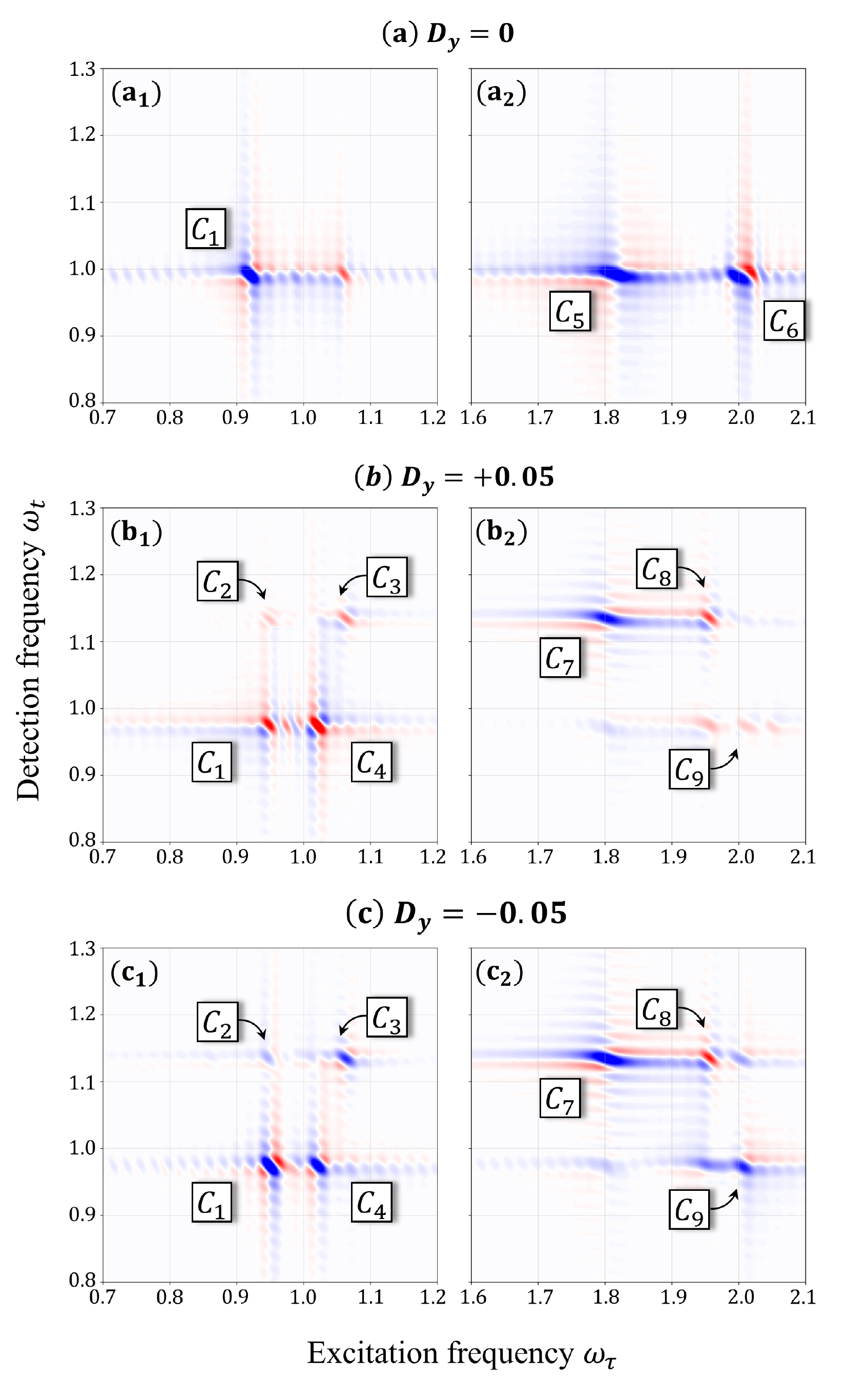}
\caption{2D THz-MR spectrum calculated from $\chi_{3}^{(2)}(\omega_\tau, \omega_t)$ for three values of $D_{y}$: (a)~$0$, (b)~$+0.05$, and (c)~$-0.05$. The intensity of each spectrum is normalized with respect to its maximum peak intensity. The red and blue areas represent positive and negative intensities. All the parameter values are the same as those in Fig.~\ref{fig.2D_t0t}.}
\label{fig.2D_0tt}
\end{figure}

We next depict $\chi_{3}^{(2)}(\omega_\tau, \omega_t)$ in Fig.~\ref{fig.2D_0tt}. Double-sided Feynman diagrams corresponding to each labeled peak are presented in Figs.~\ref{fig.2D_paths}(b) and \ref{fig.2D_paths}(c).

Peaks $C_{1}$--$C_{4}$ appear in the case $D_{y} \ne 0$ because they involve the states classified in (II), as depicted in Fig.~\ref{fig.2D_paths}. Moreover, because peaks $C_{1}$--$C_{4}$ involve the $|1_{2}\rangle$ and $|2_{3}\rangle$ states classified in (III) in each diagram in Fig.~\ref{fig.2D_paths}(b), the sign of the peak intensity changes in correspondence with the sign of $D_{y}$. Note that the peak profiles of $C_{1}$ for $D_{y}=0$ and $D_{y}>0$ are identical because the $|1_{2}\rangle$ state does not vanish even for $D_{y}=0$. We also find that the positive peak near $(\omega_\tau, \omega_t) = (1.05, 1)$ in Fig.~\ref{fig.2D_0tt}(a$_1$) arises as the consequence of the heat-bath-induced coherence $|0_0\rangle \langle 1_0|$.

In Figs.~\ref{fig.2D_paths}(b$_2$) and \ref{fig.2D_paths}(c$_2$), we observe that peaks $C_{5}$--$C_{9}$ arise from the 2Q coherence. Although the $C_{9}$ diagram involves the $|2_3\rangle$ state in (III), it overlaps with $C_{7}$ for $D_{y} \ne 0$, and the signs of the peak intensities cannot be easily evaluated. Note that here we chose a small system ($L=6$) with periodic boundary conditions, so energy states higher than the second excited state are lowered, and the profiles of peaks $C_{5}$--$C_{9}$ are distorted. Hence, within the calculations of the present model, it is more reliable to use the profiles of $C_{1}$--$C_{4}$ to determine the sign of $D_{y}$.

\section{Conclusion}
\label{sec.conclude}
A recently developed 2D THz-MR spectroscopic technique has created new possibilities for measuring complex molecular magnetic systems. In the present work, we have illustrated the key features of this technique based on nonlinear response-function theory and have described a method for simulating 2D THz-MR spectra through the use of the HEOM formalism. Using simulated 1D and 2D THz-MR spectra for a chiral magnetic material, we demonstrated that the 2D technique allows us to evaluate the pitch and direction of chirality determined from the strength and sign of the DMI, while only the absolute amplitude of the DMI can be determined from 1D measurements.

The reason the sign of the DMI was detected by 2D spectroscopy in this study is that there are eigenstates in which the phase of the wavefunction changes following the sign of the DMI, which is categorized as (III). Although there have been studies of eigenstates of spin systems with the DMI, \cite{Sakai2000} the existence of eigenstates that change phase with the sign of the DMI, as found in this study, has not been explored. It is important to know the causes of these eigenstates because they may lead to the appearance of novel phenomena.

To make a direct comparison between the results of our simulations and those obtained experimentally, however, we must increase the number of spins in accordance with those available in experimental systems. In investigating spin dynamics in the condensed phase, it is also important to consider the non-perturbative and non-Markovian system--bath interactions. Nevertheless, we believe that the present results elucidate the key features of 2D THz-MR spectroscopic methods with regard to probing the fundamental nature of a magnetic spin system.

For further investigations to monitor the ultrafast dynamical aspects of the spin system, such as the dynamics of spin waves, it is necessary to conduct a variety of advanced nonlinear spectroscopic approaches, such as pump-probe and transient absorption measurements to foster the development of this spectroscopic method. We leave such extensions to future studies, depending on progress in experimental and simulation techniques.

\section*{Acknowledgments}
The authors are thankful to Professor Jun-ichiro Kishine and Professor Keisuke Tominaga for helpful discussions. Y.T. was supported by JSPS KAKENHI (Grant No. B21H01884). J.Z. was supported by JST SPRING (Grant No. JPMJSP2110).

\subsection*{Conflict of Interest}
The authors have no conflicts to disclose.

\section*{Data availability}
The data that support the findings of this study are available from the corresponding author upon reasonable request.

\section*{REFERENCES}

\bibliography{tanimura_publist,ref_thzmr}

\end{document}